\documentclass[journal]{IEEEtran}
\usepackage{color}
\usepackage{soul}

\usepackage{cite}

\ifCLASSINFOpdf
   \usepackage[pdftex]{graphicx}
\else
\fi
\usepackage{amsmath}
\usepackage{amsmath,bm}

\begin{document}
%
\title{Discrete Huygens’ Metasurface: Realizing Anomalous Refraction and Diffraction Mode Circulation with a Robust, Broadband and Simple Design}
%
%
%

\author{Chu Qi
        and~Alex M. H. Wong,~\IEEEmembership{Senior~Member,~IEEE}
\thanks{ This work was supported by an Early Career Scheme from the Research Grants Council of the Hong Kong under Grant 21211619.(\emph{Corresponding author: Alex M. H. Wong.})}
\thanks{The authors are with the State Key Laboratory of Terahertz and Millimeter Waves, Department of Electrical Engineering, City University of Hong Kong, Hong Kong. (e-mail:chuqi2-c@my.cityu.edu.hk; alex.mh.wong@cityu.edu.hk)}%
}

%
%

\markboth{Journal of \LaTeX\ Class Files,~Vol.~14, No.~8, August~2015}%
{Shell \MakeLowercase{\textit{et al.}}: Bare Demo of IEEEtran.cls for IEEE Journals}
%



\maketitle

\begin{abstract}
Metasurfaces composed of subwavelength unit cells usually require a large number of unit cells which leads to complicated design and optimization. Aggressive discretization in a metasurface can significantly reduce the number of unit cells within a period, resulting in lower requirement for phase and/or surface impedance coverage. Additionally, the enlarged unit cells will encounter negligible mutual couplings when combined together, hence making straightforward the process of metasurface design. These advantages combine to allow the design of a novel class of metasurfaces which support the high efficiency redirection of electromagnetic (EM) waves over a wide bandwidth and operation angle. Moreover, an aggressively discretized metasurface may realize functionalities such as diffraction mode circulation, which are unsupported in its continuous counterparts. In this paper we propose a simple transmissive metasurface which can realize diffraction mode circulation by refracting plane waves with incident angles of $\bm{-}$45$\bm{^\circ}$, 0$\bm{^\circ}$, 45$\bm{^\circ}$ to plane waves with refraction angles of 0$\bm{^\circ}$, 45$\bm{^\circ}$, $\bm{-}$45$\bm{^\circ}$ respectively. The power efficiency of each anomalous refraction is more than 80\% at the design frequency of 28 GHz, and the 3-dB power efficiency bandwidth is 11\%. We fabricated and measured the metasurface, the experiment results agree well with simulation results.
\end{abstract}


%
\IEEEpeerreviewmaketitle

\section{Introduction}
%
%
%
%
\IEEEPARstart{M}{etasurfaces} are planar structures composed of subwavelength elements. They can realize various EM wave manipulation functionalities, including wave refraction, wave reflection, polarization operation, antenna beam forming, etc\cite{1,2,3,4,5,6,7,8,9,10,11,12,13,14,15}. In recent years, Huygens’ metasurfaces have attracted increasing attention because of their simultaneous electric and magnetic responses to an incident EM wave\cite{16,17,18,19,20,21,22,23,24,25,26,27}. This allows the metasurface to suppress spurious scattering and improve, among other things, the wave redirection power efficiency\cite{10,18,20,23,28}. The versatility of Huygens’ metasurfaces has been demonstrated through a variety of applications.

The general design procedure for a Huygens’ metasurface is to first design a continuous function of surface impedance (or susceptibility) which satisfies the boundary condition of the given incident EM wave and the desired reflected/transmitted EM wave. Then the metasurface is finely discretized into small elements, each of which realize the sampled values of surface impedance at their respective locations. This is done by designing the unit cells with different geometrical structures and/or parameters. Finally, the elements, which each achieve the required surface impedances at their respective locations, are fitted together to form the composite metasurface. It is assumed that when the metasurface elements are sufficiently small, the composite metasurface will behave similar to a surface whose surface impedance continuously varies in space. The detailed design process is described, for some examples, in~\cite{16,19,24,29,30,31}. Three important drawbacks manifest for a finely discretized design. Firstly, the unit cell size is electrically small (typically about one-tenth of a wavelength), increasing the fabrication difficulty. Secondly, mutual coupling among the unit cells can be significant, which can affect the performance of the metasurface. To mitigate the effect of mutual coupling, the final design often requires further optimization, which further complicates the design process. Thirdly, many different unit cells are needed to achieve the different surface impedances in a finely discretized metasurface. The need to design many different cells slows down the design process, and sometimes lands onto “blind-spots” where a given metasurface architecture cannot produce the required surface impedance. One can mitigate this either by using multiple cell architectures\cite{31}, which complicates the design, or by implementing cells with slightly different properties\cite{16}, which compromises the metasurface performance. Besides the aforementioned "blind-spots", some values of surface impedance may require strongly resonant metasurface element designs, the usage of which will compromise the robustness and bandwidth performance of the metasurface.

More recently, researchers have proposed the concept of the metagrating, which is a metasurface where each period is designed as a single scatterer, without further discretization~\cite{32}. The period of a metagrating controls the number of propagating modes. In the case of reflection with only two propagating diffraction modes (upon oblique incidence), with the specular mode suppressed, all the reflected power will couple to the desired mode, realizing high efficient anomalous reflection, sometimes with simplified structures\cite{32,33,34}. However, more complicated designs are required to manage the diffraction of more propagating modes. Some researchers introduce more degrees of freedom with extra inclusions in each period\cite{32,35,36,37}, which complicates the theory and structure, as the mutual couplings among the inclusions are difficult to rigorously analyze. Some researchers use iterative optimization methods to design metagratings based on multimode geometrical structures\cite{38,39}, which needs a lot of simulation iterations, and the optimized performance may differ depending on the specific application and the optimization algorithm. 

In parallel, our group has presented aggressively discretized metasurfaces, which is designed to have as few elements as possible within a period, and can realize efficient control of propagating diffraction modes\cite{40,41,42}. Different from metagratings which treat and design the metasurface period as a single entity, in an aggressively discretized metasurface design, we adopt the procedure of discretizing the metasurface period, and design each element separately and combine them to form the metasurface. Additionally, the aggressive discretization is not realized by simply sampling coarsely the phase/impedance profile of a continuous counterpart. Instead, we choose an appropriate discretization level from spectral domain considerations, and thereafter, we directly solve for the required transmission properties of the discretized elements. We shall show that the aggressive discretization can dramatically reduce the number of elements per period, lowering the phase/impedance coverage requirement and allowing for larger element sizes. This in turn benefits the metasurface design with simplicity and robust performance. Besides, an aggressively discretized metasurface may realize some interesting functionalities which cannot be realized by its finely discretized counterparts. For example, a metasurface with proper discretization level can realize diffraction mode circulation.

In this work, we propose an aggressively discretized metasurface with three elements per period, which can realize efficient anomalous refraction by deflecting a normal incidence to 45$^\circ$. Curiously, due to the discretization, the metasurface can also efficiently bend incident waves from $-$45$^\circ$ and 45$^\circ$ to refracted waves of 0$^\circ$ and $-$45$^\circ$ respectively, hence achieving mode circulation --- which property is not shared by the continuous or finely discretized metasurface designs. For each anomalous refraction case, our proposed metasurface can realize a power efficiency of higher than 80\% at the design frequency of 28 GHz, and a 3-dB power efficiency over a bandwidth of 11\%. Additionally, the aggressive discretization allows us to use a very simple geometrical structure, which potentially leads to a simple design with large feature sizes. This largely relaxed fabrication tolerance can further benefit high frequency even optical applications.

The rest of the paper is organized as follows. Section II introduces the metasurface discretization and design formulation. Section III details our simulation setup and results. Section IV describes our fabrication and experimental results. We discuss salient points of our work in Section V and conclude in Section VI.

\begin{figure}[tbp]
\centering
\includegraphics[width=3.4in]{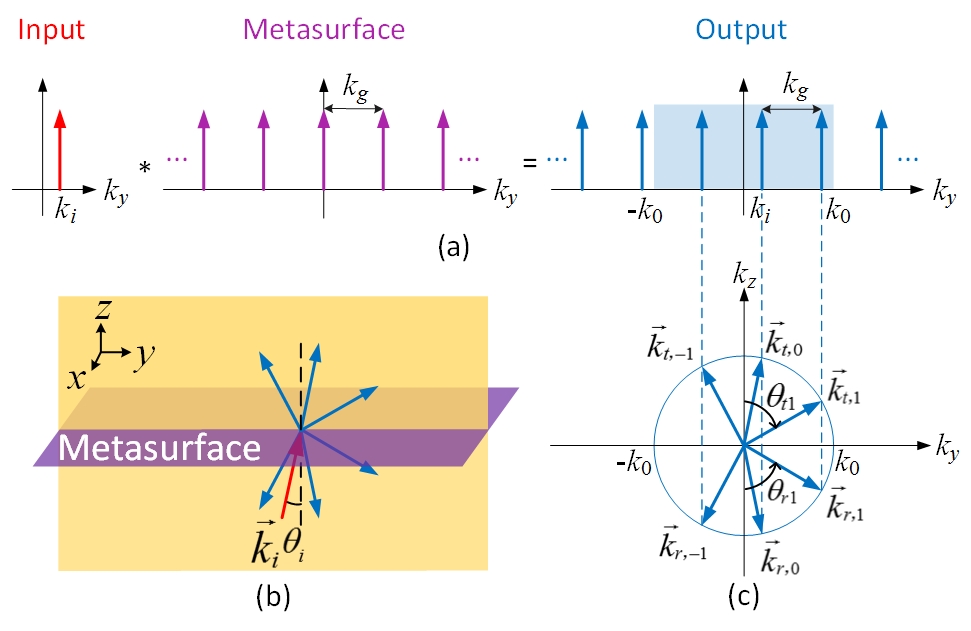}
\caption{(a) $k$-space operation of a periodic metasurface which varies along y-direction. The asterisk symbol ($\ast$) denotes the convolution operation. Arrows indicate the existence of diffraction modes with different tangential wave numbers. The blue box in the output $k$-space spectrum denotes the propagation range of $k_y\in[-k_0,k_0]$. (b) A schematic diagram of a transmissive metasurface upon an incidence with wave number $\vec k_i$ and incident angle $\theta_i$. (c) A schematic of output spectrum, $\vec k_{t,n}$ and $\vec k_{r,n}$  ($n$ = $-$1,0,1) are the wave numbers of $n$-th diffraction modes in the transmission and reflection region respectively.}
\label{KSpaceOperation}
\end{figure}

\section{{Metasurface Discretization and Design Formulation}}

In this section, we will introduce the theoretical analysis and design formulation of a discretized periodic metasurface. After that, we propose an example to realize anomalous refraction with an aggressively discretized metasurface. Additionally, we investigate the mode circulation effect in the discretized metasurface designs.

\subsection{$k$-space Operation of a Periodic Metasurface}

We find it instructive to investigate the discretization of a periodic metasurface by examining the waveform contents in the frequency domain ($k$-space). Fig.~\ref{KSpaceOperation}(a) shows the $k$-space operation of a periodic metasurface. When a periodic metasurface is illuminated by a plane wave in free space, the output (i.e. the transmitted and/or reflected wave) will consist of an infinite number of diffraction modes. The tangential wave number of $n^{\textrm{th}}$ diffraction mode in the output $k$-space spectrum is
\begin{equation}
    k_n=k_i+nk_g,
\end{equation}
where
\begin{equation}
    k_i=k_0\sin\theta_i \mbox{ and } k_g=\frac{2\pi}{\Lambda_g}.
\end{equation}
Here $k_i$ is the tangential wave number of the incident plane wave with incident angle $\theta_i$, $k_0$ is the free space wave number, $k_g$ is the wave number of the metasurface with period $\Lambda_g$. Of the infinite number of diffraction modes, only the ones in the propagation range of $k_y\in[-k_0,k_0]$ can scatter into the far field, and the ones out of the propagation range are evanescent. Fig.~\ref{KSpaceOperation}(b) is a schematic diagram of a transmissive metasurface in Fig.~\ref{KSpaceOperation}(a). In a transmissive metasurface, there is a same number of propagating diffraction modes ($n$ = -1, 0, 1) in transmission and reflection spectrum respectively. The refraction/reflection angles can be determined using

\begin{equation}
    \theta_{t,n}=\theta_{r,n}=\sin^{-1}\frac{k_n}{k_0}.
\end{equation}
Fig.~\ref{KSpaceOperation}(c) is a schematic of the reflected and transmitted waves, corresponding to the output spectrum shown in Fig.~\ref{KSpaceOperation}(a).

While the tangential wave number of the incident wave ($k_i$) and metasurface period ($\Lambda_g$) determine the tangential wave numbers of the diffraction modes ($k_n$) in the output spectrum, the structure of the metasurface and its interaction with the incident EM wave will determine the magnitude and phase of each diffraction mode. That is, by engineering the transmission and/or reflection characteristics of the discretized metasurface elements, we can engineer the transmission and/or reflected diffraction modes. In the following we describe a way to calculate the required transmission coefficients for discrete set of metasurface elements, which will transform a known incident wave into a a desired set of transmitted plane waves.

\subsection {Discretized Metasurface --- Design Formulation}

\begin{figure}[tbp]
\centering
\includegraphics[width=3.2in]{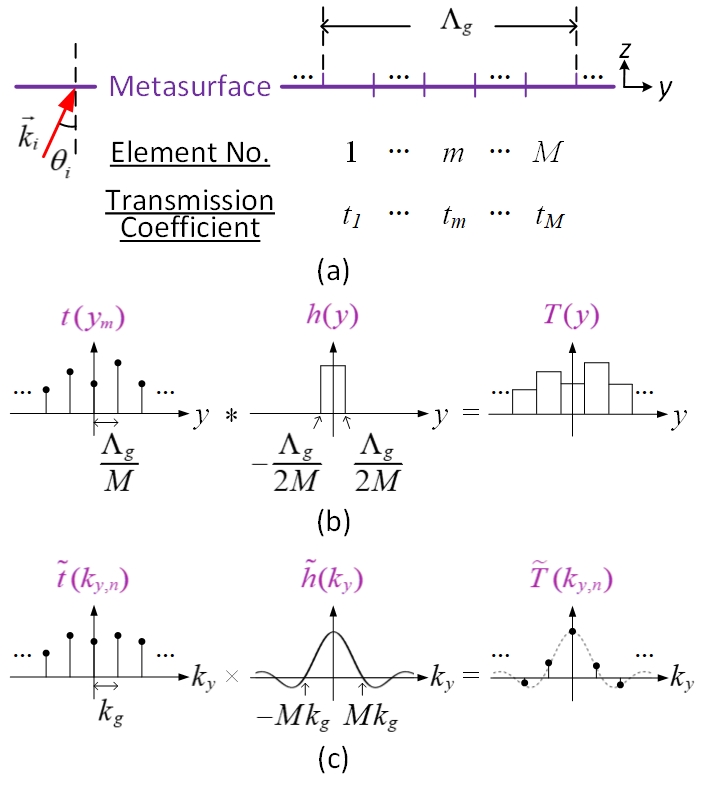}
\caption{(a) Schematic of one period of a transmissive metasurface with a discretization level of $M$ elements per period. (b) Transmission of the metasurface in the spatial (y) domain. (c) Transmission of the metasurface in the tangential wave number ($k_y$) domain. The asterisk operator ($\ast$) denotes the convolution operation and the multiplication sign ($\times$) denotes the multiplication operation.}\label{R_DiffractionModeEngineering}
\end{figure}

Consider a transmissive metasurface illuminated by a plane wave with tangential wave number $k_i$, we can write the transmitted electric field at $z=0^+$:

{\begin{equation}     E_t(y)=E_ie^{-jk_iy}T(y), \end{equation}}

\noindent where $E_i$ is a constant representing the complex amplitude of the incident electric field. $T(y)$ is the transmission function of the metasurface along the variation direction (y-direction). Fig.~\ref{R_DiffractionModeEngineering}(a) shows a schematic of a metasurface with period $\Lambda_g$ and a discretization level of $M$ elements per period: each element has a size of $\frac{\Lambda_g}{M}$ in the variation direction, the $m^{\textrm{th}}$ element has a transmission coefficient of $t_m$, which can be written as

{\begin{equation} t_m=|t_m|e^{j\varphi_m}, \end{equation}}

\noindent where $|t_m|$ and $\varphi_m$ are the magnitude and phase of the transmission coefficient associated with the element. Figs.~\ref{R_DiffractionModeEngineering} (b) and (c) show the transmission of the periodic metasurface in the spatial (y) domain and the tangential wave number ($k_y$) domain respectively. As shown in Fig.~\ref{R_DiffractionModeEngineering}(b), $T(y)$ is the transmission function of the discretized metasurface, where $t(y_m)$ and $h(y)$ are the array function and element function of the transmission coefficients respectively. That is

{\begin{equation}    T(y)=t(y_m)\ast h(y), \end{equation}}

\noindent where $t(y_m )=\sum_m t[y_m ]\delta(y-y_m)$, $t[y_m ]$ is the transmission coefficient of the element located at $y_m$. Defining $\Tilde{t}(k_{y,n})$, $\Tilde{h}(k_{y})$ and $\Tilde{T}(k_{y,n})$ as the Fourier transform pairs of $t(y_m)$, $h(y)$ and $T(y)$ respectively, we have

{\begin{equation}    \Tilde{T}(k_{y,n})=\Tilde{t}(k_{y,n})\times\Tilde{h}(k_y). \end{equation}}

\noindent From Fourier transform theory, $\Tilde{T}(k_{y,n})$ can be written as a summation of delta functions where each delta function corresponds to a diffraction mode in the transmission spectrum, i.e., $\Tilde{T}(k_{y,n})=\sum_n\Tilde{T}[k_{y,n}]\delta(k_y-k_{y,n})$. Since $t[y_m ]$ is discrete with spacing $\frac{\Lambda_g}{M}$ between adjacent elements and periodic of $M$ elements, Fourier transform theory guarantees that, $\Tilde{t}[k_{y,n}]$ will be discrete with interval $k_g=\frac{2\pi}{\Lambda_g}$ between adjacent modes and periodic of $M$ modes. Using $t[m]$ and  $\Tilde{t}[n]$ to denote the $m^{\textrm{th}}$ and $n^{\textrm{th}}$ items in $t[y_m]$ and $\Tilde{t}[k_{y,n}]$ respectively, we have

{\begin{equation} \begin{aligned} & t[m]=\sum_{n}\Tilde{t}[n]e^{jn\frac{2\pi}{M}m} \\ & \Tilde{t}[n]=\sum_{m=1}^{M}t[m]e^{-jn\frac{2\pi}{M}m}, \end{aligned} \end{equation}}

\noindent which is to say, $t[m]$ and $\Tilde{t}[n]$ are Fourier series pairs. Additionally,

{\begin{equation} \begin{aligned}    & h(y)=    \begin{cases}    1,\quad &-\frac{\Lambda_g}{2M}<y<\frac{\Lambda_g}{2M} \\    0,\quad &otherwise     \end{cases} \\    & \Tilde{h}(k_y)=\frac{2\pi}{Mk_g}\mbox{sinc}(\frac{k_y}{Mk_g}).\end{aligned}\end{equation}}

\noindent The $n^{\textrm{th}}$ item of $\Tilde{T}[k_{y,n}]$, which we denote as $\Tilde{T}[n]$, corresponds to the coefficient of the $-n^{\textrm{th}}$ diffraction mode in the transmission spectrum. That is, we can find the coefficient of the $n^{\textrm{th}}$ diffraction mode, $a_n$, using (8)--(9):

{\begin{equation}    \begin{aligned}  a_n=E_i\Tilde{T}[-n]  & =E_i\Tilde{t}[-n]\times\Tilde{h}(-nk_g) \\  & =\frac{E_i\Lambda_g}{M^2}\mbox{sinc}(\frac{-n}{M})\sum_{m=1}^{M}t[m]e^{jn\frac{2\pi}{M}m}.   \end{aligned} \end{equation}}

\noindent Therefore, with a given discretization level ($M$) and transmission coefficient of each element ($|t_m|$ and $\varphi_m$ for $m=1,\cdots,M$), we can calculate the magnitude ($|a_n|$) and phase ($\angle a_n$) of each diffraction mode in the transmission spectrum. Here we consider the element function $h(y)$ as a rectangular function, which means we assume wave-transmission homogeneity within each element. Besides, we assume the invariance of the element function $h(y)$ with different $k_i$, which means the element function remains the same for different incidence angles. The metasurface may be slightly anisotropic, but such spatial dispersion effect is minor because in our case the element size ranges typically between quarter to half wavelength, which is still small compared to the wavelength. When the incidence and transmission angles are within a relatively small range, the change in $h(y)$ with respect to \{$\theta_i, \theta_t$\} will be gradual. Nevertheless, one can always accurately account for spatial dispersion by incorporating the accurate $\Tilde{h}(k_y)$ corresponding to the element function $h(y)$.

While the preceding formulation analyzes the diffraction modes excited with a given set of transmission coefficients, we now present a formulation to find the required transmission coefficients which synthesize a desired transmission spectrum. Writing (10) in matrix form, we have

{\begin{equation}    \vec a=C_0D\vec t, \end{equation}}

\noindent where

{\begin{equation} \begin{aligned} &\vec{a}=\begin{bmatrix} \vdots \\\ a_{-1} \\\ a_0 \\\ a_1 \\\ \vdots \end{bmatrix}_{N\times1}, \vec{t}=\begin{bmatrix}t_1 \\\ \vdots \\\ t_M \end{bmatrix}_{M\times1}, C_0=\frac{E_i\Lambda_g}{M^2}, \\ &D=\begin{bmatrix} \vdots &\ddots &\vdots \\\ \mbox{sinc}(\frac{-1}{M})e^{-j2\pi\frac{1}{M}} &\cdots &\mbox{sinc}(\frac{-1}{M})e^{-j2\pi\frac{M}{M}} \\\ 1 &\cdots &1 \\\ \mbox{sinc}(\frac{1}{M})e^{j2\pi\frac{1}{M}} &\cdots &\mbox{sinc}(\frac{1}{M})e^{j2\pi\frac{M}{M}} \\\ \vdots &\ddots &\vdots \end{bmatrix}_{N\times M}.\end{aligned}\end{equation}}

\noindent $\vec a$ is an $N$-element vector composing of the coefficients of the diffraction modes which need to be controlled. When $M\geq N$, (11)-(12) can be solved to find the required elements' transmission coefficients ($\vec t$) for a given transmission spectrum ($\vec a$). When $M=N$, the solution is unique. That is, to realize control of $N$ diffraction modes, the lowest number of elements required (the most aggressive discretization level) is $M=N$.



\begin{figure}[tbp]
\centering
\includegraphics[width=3.2in]{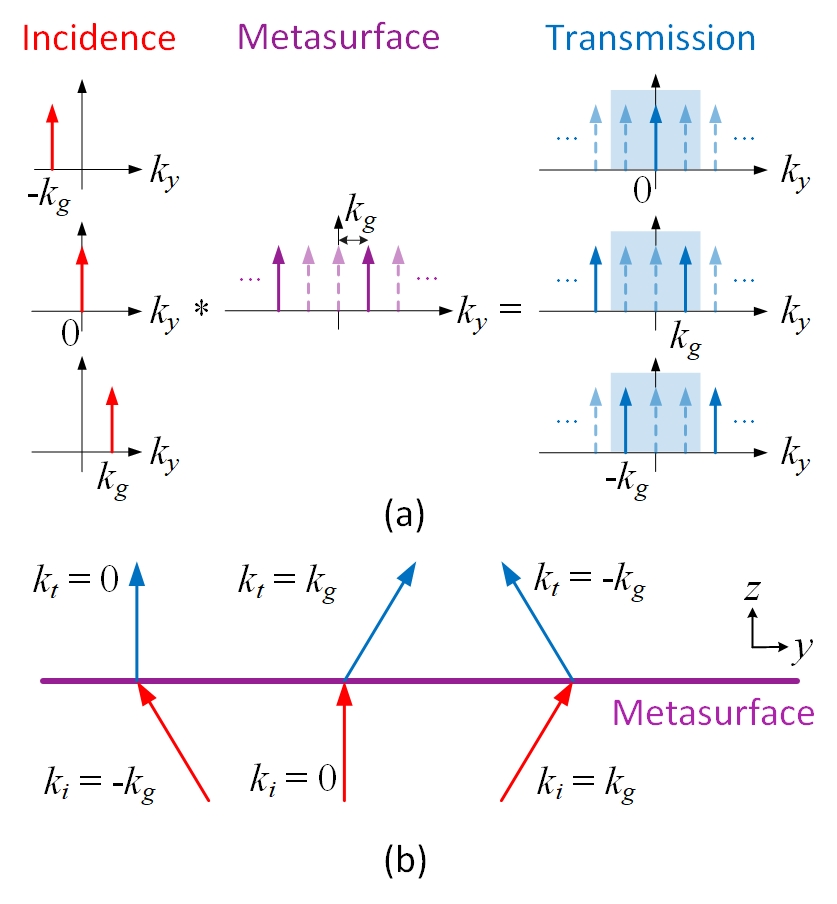}
\caption{(a) The $k$-space operation of a metasurface featuring the circulation of three diffraction modes. Arrows with solid line represent diffraction modes with non-zero magnitude, while arrows with dashed line represent diffraction modes with zero magnitude. (b) A schematic diagram of the transmissive metasurface in (a).}
\label{kSpaceOperation_3 modes}
\end{figure}

\subsection{Anomalous Refraction Metasurface Design}

We now use the aforementioned method to design an aggressively discretized metasurface which realizes anomalous refraction upon a normal incidence. To realize perfect anomalous refraction, the metasurface should be designed to ensure that for the propagating diffraction modes in the refraction spectrum, only one (the desired) diffraction mode carries power. Here we consider the case of three propagating diffraction modes, i.e., $k_g\in(\frac{k_0}{2},k_0)$. That is, we concentrate on the scenario where one anomalously deflects a normal incidence by more than 30$^\circ$. Applying (1), one can see that the metasurface has three propagating diffraction modes. Further, only the $1^\textrm{st}$ diffraction mode (placed at $k_g$) has non-zero magnitude. The three propagating diffraction modes can be controlled with an aggressive discretization level of $M=3$. We can solve for the three elements' transmission coefficients using (11)--(12) with the condition $[a_{-1}, a_0, a_1]=[0, 0, a_1]$. The resultant transmission coefficient, normalized to the first term, is $[1, e^{-j\frac{2\pi}{3}}, e^{j\frac{2\pi}{3}}]$. In other words, the three elements should have uniform transmission magnitudes and equidistant transmission phases. Fig.~\ref{kSpaceOperation_3 modes} shows the $k$-space operation and the refraction property of this metasurface (the case with $k_i=0$); in the proceeding sections we will show simulation and experimental results demonstrating the achievement of anomalous refraction as depicted hereby.

Previously we have demonstrated that a metasurface with an aggressive discretization level of three elements per period can realize efficient anomalous refraction of a normal incidence~\cite{42}. The design in~\cite{42} is based on a finely discretized Huygens' metasurface proposed in~\cite{16}, which has twenty unit cells per period as shown in Fig.~\ref{FabricatedMS1}(c). In~\cite{42}, we use the same unit cell geometrical structure to realize a similar anomalous refraction functionality as in~\cite{16}. We then aggressively discretize the metasurface to three elements per period where each element is composed of three identical unit cells. Due to the aggressive discretization, the number of elements (types of unit cells) per period is significantly reduced from twenty to three. Compared to~\cite{16}, the metasurface proposed in~\cite{42} achieves a slightly improved power efficiency. Here in this work, leveraging the flexibility afforded by the increased element size, we investigate what happens to the feature size, power efficiency and operation bandwidth of a metasurface where each element is implemented as a single unit cell.

\subsection{The Mode Circulation Effect}

Due to aggressive discretization in the spatial domain, the metasurface assumes a periodic behaviour in $k$-space where the $k$-domain period is smaller than those of finely discretized metasurfaces. This leads to an interesting functionality which we term the mode circulation effect. Fig.~\ref{kSpaceOperation_3 modes}(a) shows the $k$-space operation for the metasurface designed in the previous subsection. As we can see, for this metasurface, upon incidence with $k_{i}=-k_g$, the only non-zero diffraction mode within the propagation range in the transmission $k$-space spectrum is $k_{t,1}=k_{i}+k_g=0$; upon incidence with $k_{i}=0$, the only non-zero diffraction mode within the propagation range in the transmission $k$-space spectrum is $k_{t,1}=k_{i}+k_g=k_g$; upon incidence with $k_{i}=k_g$, the only non-zero diffraction mode within the propagation range in the transmission $k$-space spectrum is $k_{t,-2}=k_{i}-2k_g=-k_g$. That is, a metasurface with $k_g\in(\frac{k_0}{2},k_0)$ and a discretization level of three elements per period can realize the circulation of three diffraction modes with tangential wave numbers of $-k_g$, 0 and $k_g$. Fig.~\ref{kSpaceOperation_3 modes}(b) is a schematic of the corresponding metasurface featuring diffraction mode circulation. 

\begin{figure}[tbp]
\centering
\includegraphics[width=3.2in]{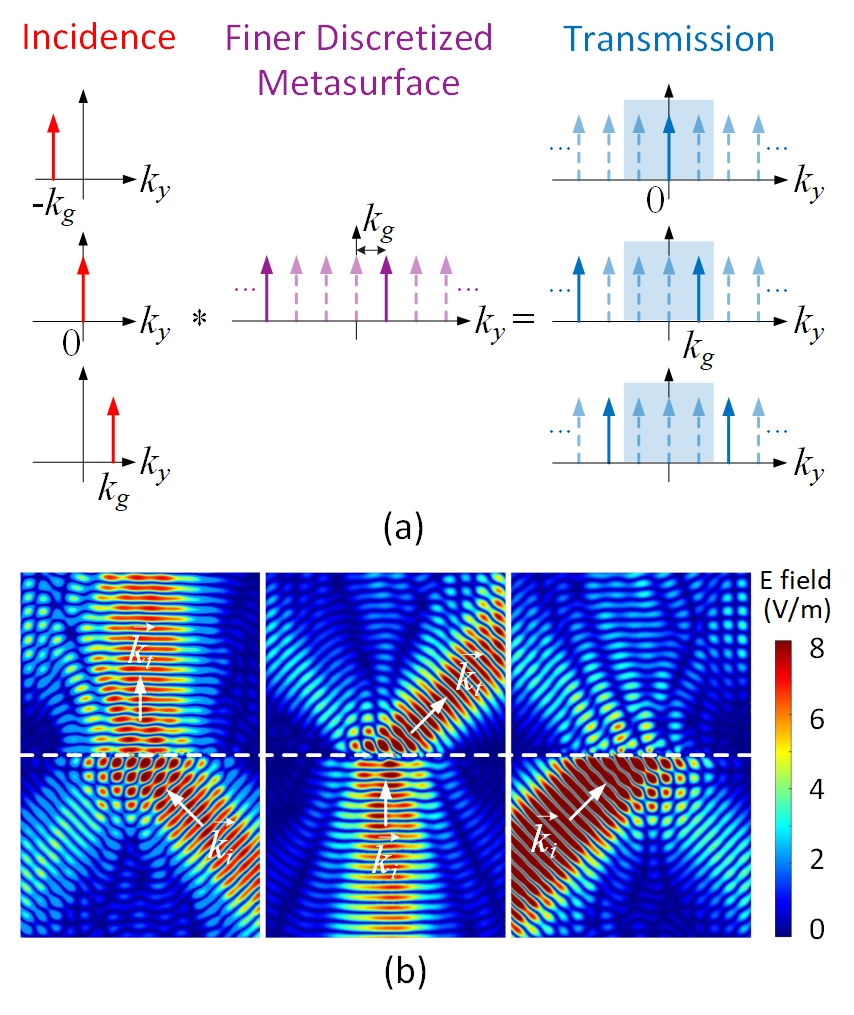}

{\caption{(a) The $k$-space operation of a metasurface with the same $k_g$ as the metasurface in Fig.~\ref{kSpaceOperation_3 modes}, but with a finer discretization level of four elements per period. (b) Simulated electric field magnitude distributions of the metasurface in (a) upon Gaussian beams with incident angles of $-$45$^\circ$, 0$^\circ$ and 45$^\circ$ (Front view, dashed line: location of the metasurface).}\label{Discretization4}}
\end{figure}

\begin{figure*}[ht]
\centering
\includegraphics[width=6.2in]{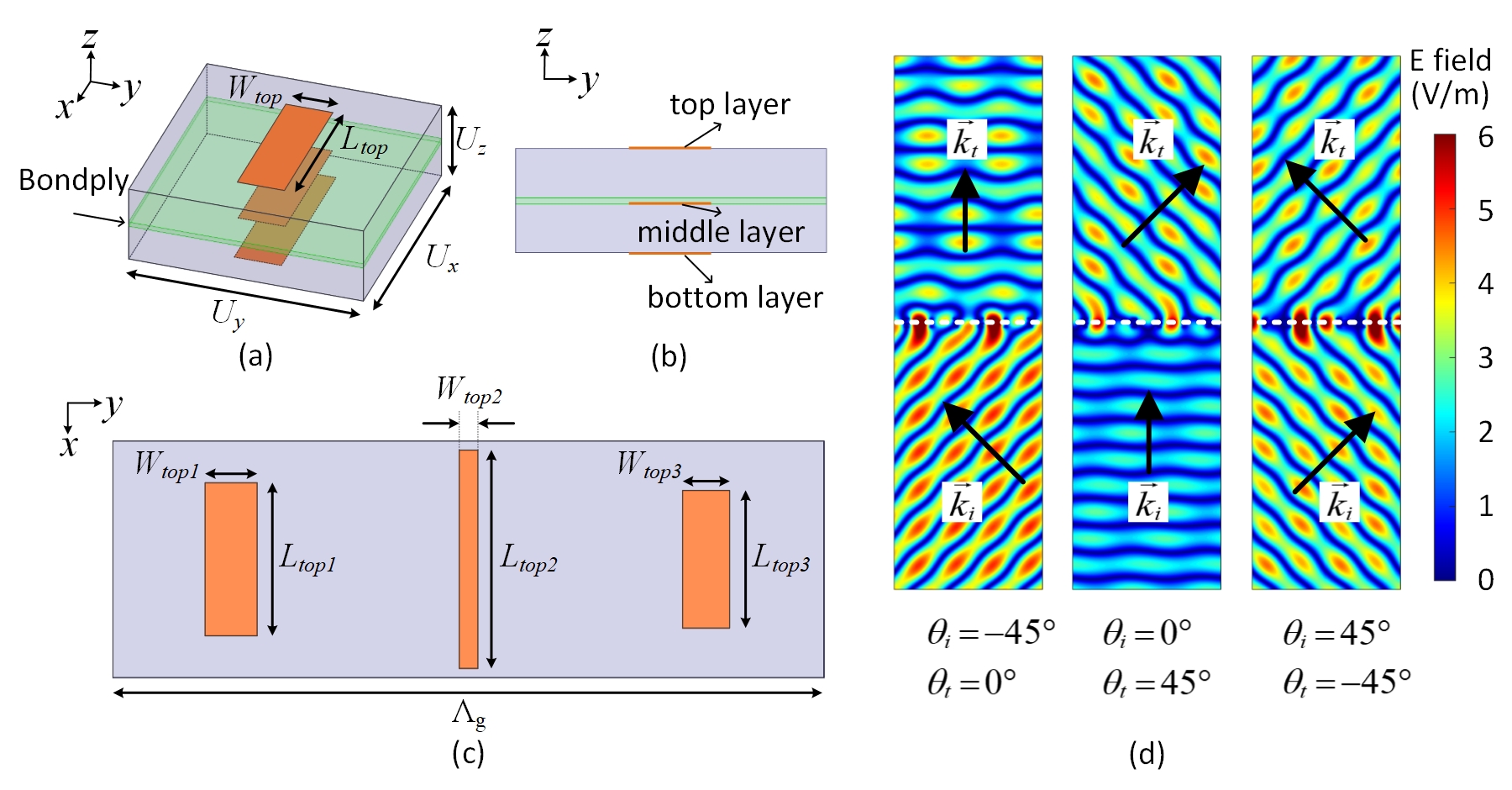}
\caption{(a) Perspective view and (b) Front view of the geometrical structure of the unit cell. The size of the unit cell is $U_x=U_y$ = 5 mm, $U_z$ = 1.67 mm.  (c) Top view of one period of the proposed metasurface. (d) The electric field magnitude distributions of the metasurface upon incident plane waves with incident angles of $-$45$^\circ$, 0$^\circ$ and 45$^\circ$ respectively.}
\label{structure1}
\end{figure*}

The mode circulation effect hereby presented cannot be achieved unless careful consideration is given based on the $k$-space operation. As we see, the realization of mode circulation phenomenon involves some diffraction modes which are evanescent in the case of normal incidence. These \emph{evanescent} diffraction modes are often implicitly set to zero in the design of finely discretized metasurfaces, which are often based on the fine sampling of the continuous impedance/surface current profile, obtained as a surface boundary that transforms an incident plane wave into a \emph{propagating} output wave. Therefore, a phenomenon like mode circulation is highly unobvious in the design of finely discretized metasurfaces. For an example, Fig.~\ref{Discretization4} shows a finer discretized metasurface with four elements per period which realizes the same anomalous refraction (upon normal incidence) as in previous subsection. That is, the metasurface has the same $k_g$ as in Fig.~\ref{kSpaceOperation_3 modes}(a), but with a finer discretization of $M=4$. This discretization level leads to a period of $4k_g$ in the $k_y$ domain of the metasurface: beside the $1^{\textrm{st}}$ diffraction mode, the next non-zero diffraction mode of significance is the $-3^{\textrm{rd}}$ one (instead of the $-2^{\textrm{nd}}$ diffraction mode in the aggressively discretized design). Hence, for incidence with tangential wave numbers of $-k_g$ and 0, the transmission $k$-space spectrum will have a single non-zero diffraction mode within the propagation range, with tangential wave numbers of 0 and $k_g$ respectively. However, for incidence with tangential wave number of $k_g$, there will be no non-zero diffraction mode within the propagation range. This causes weak transmission and does not lead to mode circulation. As a demonstration of this phenomenon, we simulate a metasurface with the same period and unit cell structure as our proposed metasurface in section III, and change the discretization level to four elements per period. The four elements have uniform transmission magnitudes and equidistant transmission phases. Fig.~\ref{Discretization4}(b) shows the simulated electric field magnitude distributions of the metasurface upon Gaussian beams with incident angles of -45$^{\circ}$, 0$^{\circ}$ and 45$^{\circ}$, respectively. As we can see, for incident beams with tangential wave numbers of $-k_g$ and 0, the metasurface can somewhat efficiently refract the beams to diffraction modes with tangential wave numbers of 0 and $k_g$. However, for an incident beam with a tangential wave number of $k_g$, most of the incident power is reflected, which is consistent with the $k$-space operation showing no non-zero diffraction mode within the propagation range of the transmission spectrum.

In this section, we have theoretically shown, based on the metasurface discretization and design formulation, that a periodic metasurface can realize efficient anomalous refraction of a normal incidence with an aggressive discretization level of three elements per period. Additionally, the aggressive discretization leads to mode circulation effect, which is highly unobvious and heretofore unreported for finely discretized metasurfaces. In the following sections, we will report simulation and experimental results showing the realization of anomalous refraction and mode circulation by a discrete metasurface with high efficiency and broadband performance.

\section{Metasurface Design and Simulation}

In this section, we report the design of a transmissive discrete Huygens’ metasurface, as an example to show that a discrete metasurface can achieve anomalous refraction and mode circulation with good performance. We design a metasurface with $k_g=\frac{k_0}{\sqrt{2}}$. Equations (1)--(3) show that this metasurface can realize mode circulation by refracting incident waves with incident angles of -45$^{\circ}$, 0$^{\circ}$, 45$^{\circ}$ to transmitted waves with refraction angles of 0$^{\circ}$, 45$^{\circ}$, -45$^{\circ}$ respectively. The period of this metasurface is
\begin{equation}
    \Lambda_g=\frac{2\pi}{k_g}=\sqrt{2}\lambda_0,
\end{equation}
where $\lambda_0$ is the free space wavelength. We design the metasurface with $\Lambda_g=15$ mm at 28 GHz, the metasurface is aggressively discretized to three elements per period, and we design each element as one unit cell. Ansys HFSS is used for the simulation. Figs.~\ref{structure1}(a) and (b) show the geometrical structure of the unit cell, which is constructed by three layers of rectangular patches. It is designed on two Rogers RT/duroid 5880 boards ($\epsilon_r$ = 2.2, $\delta_t$ = 0.0009) with thickness of 0.787 mm and 17.8 $\mu$m (1/2 oz.) copper cladding bonded by a bondply Rogers RO 4450F ($\epsilon_r$ = 3.52, $\delta_t$ = 0.004) with thickness of 0.1 mm. The electric field is along x-axis. By sweeping the widths and lengths of the three rectangular patches, we can get the transmission coefficients of the unit cell with different geometrical parameters. We choose three unit cells with approximately the same transmission magnitudes (about 0.9) and equidistant transmission phases (120$^\circ$) to compose the period of the metasurface, which is shown in Fig.~\ref{structure1}(c). The geometrical parameters are given in Table I. From 2D periodic simulation results, this metasurface can refract incident waves with incident angles of $-$45$^\circ$, 0$^\circ$ and 45$^\circ$ to transmitted waves with refracted angles of 0$^\circ$, 45$^\circ$ and $-$45$^\circ$ with power efficiencies of 81.3\%, 81.8\% and 80.1\% respectively. The simulated 3-dB power efficiency bandwidth of this metasurface is 11\%. Fig.~\ref{structure1}(d) shows the electric field magnitude distributions of the metasurface under incident plane waves with incident angles of $-$45$^\circ$, 0$^\circ$ and 45$^\circ$ respectively, from which we can see the mode circulation effect of this metasurface.

\begin{table}[ht]
    \caption{Geometrical Parameters of the Unit Cells}
    \centering
    \begin{tabular}{c|cccccc}
    \hline
        Unit cell No. & $L_{top}$ & $W_{top}$ & $L_{mid}$ & $W_{mid}$ & $L_{bot}$ & $W_{bot}$\\
        & [mm]& [mm] & [mm] & [mm] & [mm] & [mm]\\
        \hline
         1 & 3.2 & 1.1 & 3.0 & 1.3 & 3.4 & 0.9\\
         2 & 4.6 & 0.4 & 2.6 & 0.6 & 4.7 & 0.6\\
         3 & 2.9 & 1.0 & 2.4 & 1.0 & 2.9 & 1.0\\
    \hline
    \end{tabular}
\end{table}

\begin{figure*}[ht]
\centering
\includegraphics[width=5.8in]{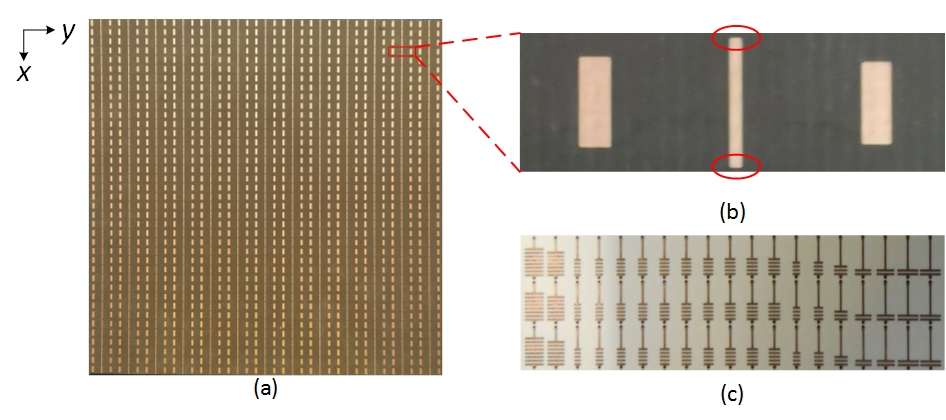}
\caption{(a) Fabricated metasurface. (b) One period of our proposed metasurface compared with (c) one period of the metasurface proposed in~\cite{16}.}
\label{FabricatedMS1}
\end{figure*}

\begin{figure*}[ht]
\centering
\includegraphics[width=6.5in]{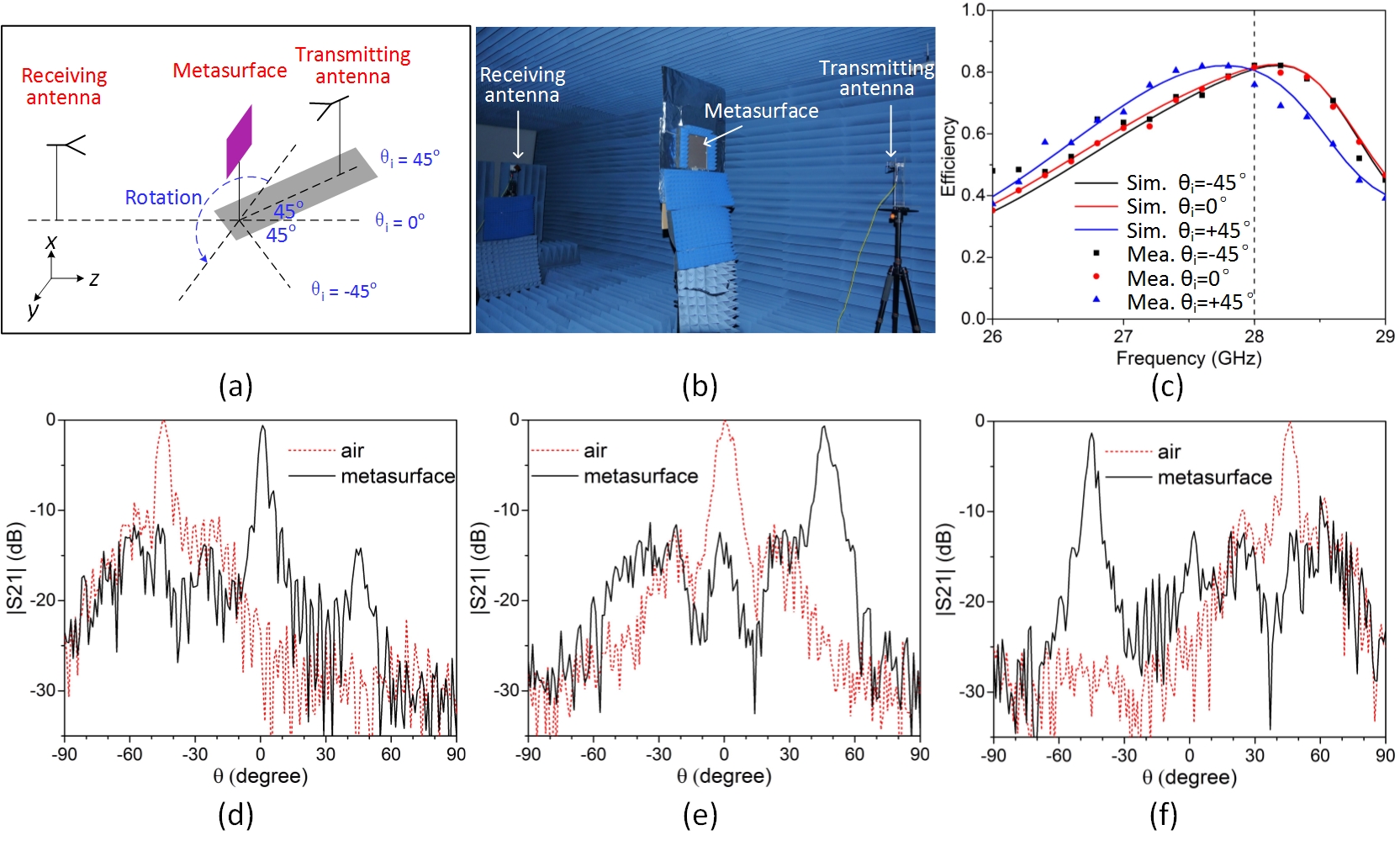}
\caption{Experiment. (a) A schematic of experimental setup. (b) A photo of the experimental setup. (c) The measured and simulated anomalous refraction power efficiencies. (d)-(f) Measured scattering of the metasurface compared with the scattering of air (same setup without metasurface) upon incidences of $-$45$^\circ$, 0$^\circ$, 45$^\circ$ respectively.}
\label{Experiment1}
\end{figure*}

\section{Fabrication and Experiment}

We proceed to fabricate and measure the proposed metasurface. Fig.~\ref{FabricatedMS1}(a) is a photo of the fabricated metasurface. As mentioned in the previous section, the metasurface is fabricated by etching two Rogers RT/duroid 5880 boards ($\epsilon_r$ = 2.2, $\delta_t$ = 0.0009) with thickness of 0.787 mm and 17.8 $\mu$m (1/2 oz.) copper cladding, then bonding the two boards using Rogers RO 4450F ($\epsilon_r$ = 3.52, $\delta_t$ = 0.004) bondply with a thickness of 0.1 mm. The size of the fabricated metasurface is 195$\times$195 mm$^2$. Figs.~\ref{FabricatedMS1}(b) and (c) show one period of our proposed metasurface and a transmissive Huygens’ metasurface proposed in~\cite{16}, which can realize anomalous refraction by deflecting a normal incidence by 30$^\circ$. We observe that, compared to finely discretized metasurface design, an aggressively discretized metasurface can lead to much larger unit cells and benefit from very simple design, which relaxes fabrication tolerances. More detailed comparison is presented in the discussion section.

\begin{figure*}[ht]
\centering
\includegraphics[width=6.5in]{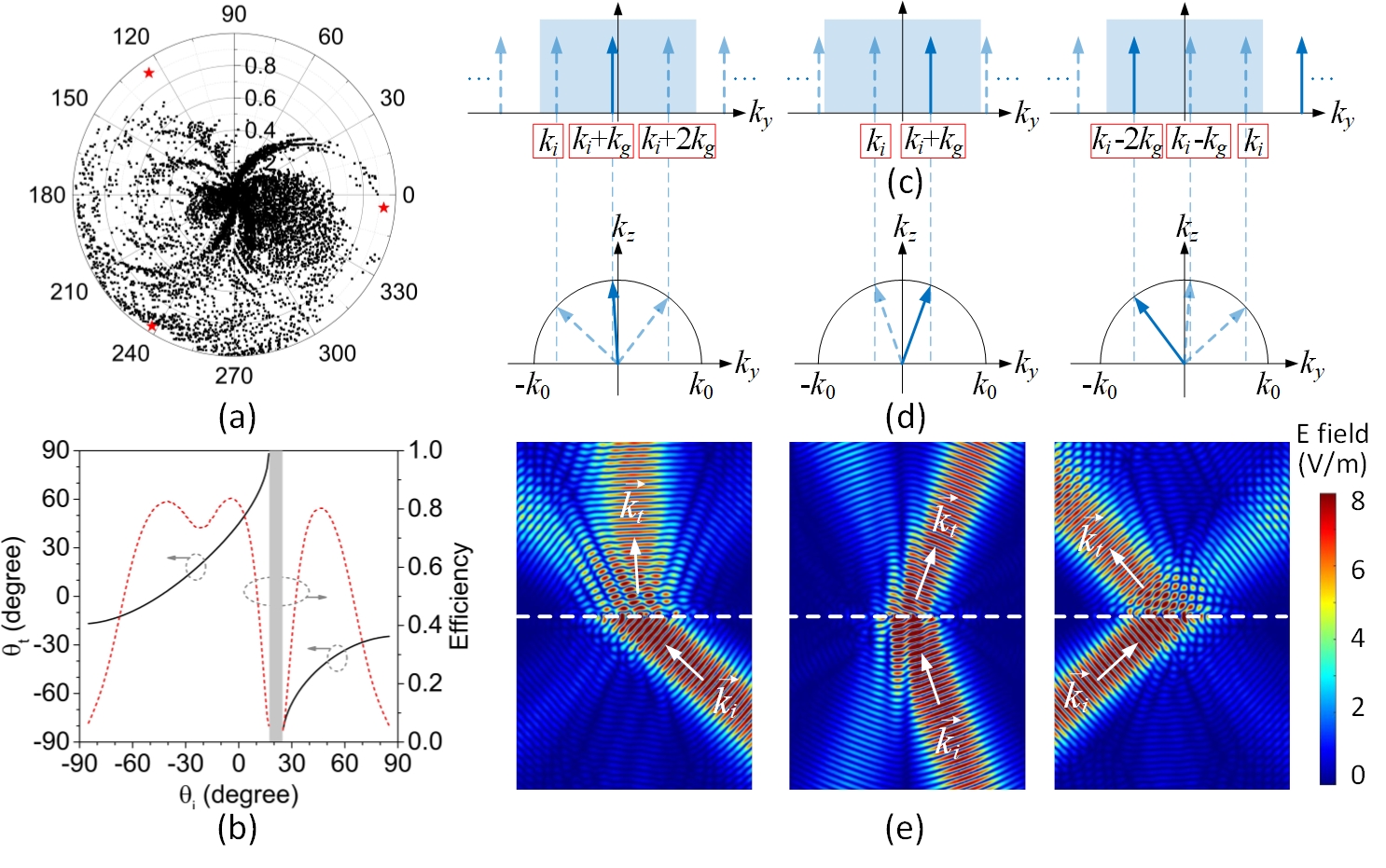}

{\caption{(a) Polar plot of transmission coefficients of our proposed unit cell structure with different geometrical parameters. Black dots are the result of parametric sweep: $L_{top}$, $L_{mid}$, $L_{bot}$ are linear swept from 2.5 mm to 4.9 mm with a step of 0.4 mm, while $W_{top}$, $W_{mid}$, $W_{bot}$ are linear swept from 0.5 mm to 1.5 mm with a step of 0.5mm. Red dots: transmission coefficients of the three unit cells in our proposed design, their geometrical parameters are shown in Table I. (b) Refraction angle and anomalous refraction power efficiency of the proposed metasurface as a function of the incidence angle. (c) The transmission $k$-space spectrums and (d) the corresponding transmission schematics of our proposed metasurface upon incidences with $\theta_i$ = $-$50$^\circ$,$-$20$^\circ$ and 50$^\circ$ respectively. (e) Simulated electric field magnitude distributions of our proposed metasurface upon Gaussian beams with incident angles of $-$50$^\circ$, $-$20$^\circ$ and 50$^\circ$.\label{Discussion}}}
\end{figure*}

Fig.~\ref{Experiment1}(a) shows a schematic of the experimental setup. In the setup, the transmitting antenna (a 4-40 GHz double ridged horn NSI-RF-RGP-40) is fixed with the metasurface with incident angles of $-$45$^\circ$, 0$^\circ$ and 45$^\circ$ respectively, for the three refraction cases. The receiving antenna (an 18-40 GHz standard diagonal horn FR-6413) rotates to measure the scattering of the metasurface at the transmission side. Fig.~\ref{Experiment1}(b) shows a photo of the experimental setup. In the experiment, the metasurface is embedded into a metallic board to eliminate direct transmission between the transmitting and receiving horns. Figs.~\ref{Experiment1}(c)-(f) show the experimental results. Fig.~\ref{Experiment1}(c) is the measured anomalous refraction power efficiencies compared with the simulation results. The measured power efficiency is the ratio between the power received in the desired refraction direction and the power illuminated on the metasurface. At the design frequency of 28 GHz, the measured power efficiencies of the fabricated metasurface are 82.2\%, 81.6\% and 76.0\% for anomalous transmission cases of incidences from $-$45$^\circ$, 0$^\circ$ and 45$^\circ$ to refractions of 0$^\circ$, 45$^\circ$ and $-$45$^\circ$ respectively. Besides, the efficiency performance over the 3-dB frequency band agrees well with the simulated result. Figs.~\ref{Experiment1}(d), (e) and (f) are the measured scattering of the metasurface compared with the scattering of air under incident angles of $-$45$^\circ$, 0$^\circ$ and 45$^\circ$ respectively. As we can see, the metasurface can efficiently refract plane waves with incident angles of $-$45$^\circ$, 0$^\circ$ and 45$^\circ$ to refracted waves with refraction angles of 0$^\circ$, 45$^\circ$ and $-$45$^\circ$ respectively, realizing diffraction mode circulation. In all the three cases, sidelobe levels are lower than -10 dB, showing that the metasurface achieves a strong suppression of spurious scattering along with high-efficiency anomalous transmission to the desired modes.

\section{Discussion}

In this section, we compare our work with some previously presented metasurface designs, including finely-discretized metasurfaces and metagratings. By comparing the feature sizes and efficiency/bandwidth performances in different works, we will show that the aggressive discretization in metasurface design can lead to simple structure with large feature size, resulting in efficient and broadband performance. Additionally, we investigate the anomalous refraction performance of our proposed metasurface with different incidences, which shows that the discrete metasurface design formulation we presented here is systematic and straightforward.

In discrete metasurface design, the enlarged unit cell size and simple structure can greatly relax the fabrication tolerance, which will benefit the metasurface design at high frequencies and even optical frequencies. As an example, we compare our proposed metasurface with a finely discretized transmissive Huygens’ metasurface proposed in~\cite{16}. The structures of the fabricated metasurfaces are shown in Figs.~\ref{FabricatedMS1}(b) and (c). Clearly the aggressively discretized metasurface feature larger elements than the finely discretized counterpart. We proceed to compare the critical feature sizes in both metasurfaces. We use the electrical length to account for the fact that the metasurfaces are built for different frequencies of operation. The minimum feature size in our design is 0.03$\lambda_0$, which is about 6-fold increased from the design in~\cite{16}. In our design, the smallest gap (0.3 mm) is easily fabricable by conventional etching technology. Extension to even higher frequency (up to 100 GHz) should be straightforward. We attempt to further increase the minimum feature size by removing the two smallest gaps. The one on the top layer is circled in Fig.~\ref{FabricatedMS1}(b); gaps of similar dimensions exist in the bottom layer. The resultant structure boasts a minimum feature size of 0.15$\lambda_0$, while the simulated anomalous refraction efficiency slightly reduced to 70\%. This amended metasurface has a similar anomalous refraction efficiency as~\cite{16}, but a minimum feature size which is 28-fold increased. Despite the relaxed feature size, our proposed metasurface produces at a larger refraction angle and operates with higher anomalous refraction power efficiency. Similar feature size improvements are observed in comparison to more recent works on bianisotropic Huygens' metasurfaces~\cite{24,30,lavigne2018susceptibility}, but a quantitative comparison is not performed with these works since the refraction angles are drastically different. 

Metagratings, with one period designed as a single scatterer, can potentially alleviate the problem of small feature size in finely discretized metasurface designs. However, the theory and design method of some existing metagratings, aiming at achieving high efficient anomalous reflection/refraction at the design frequency, may result in limited bandwidth performance. In the aggressively discretized metasurface design, one can use simple structure and choose unit cells away from the resonance, leading to robust design with broadband performance. Fig.~\ref{Discussion}(a) shows the simulated transmission coefficients of the unit cell in Fig.~\ref{structure1}(a) with different geometrical parameters. As we sweep the lengths and widths of the electric dipoles on three layers, we can see that some transmission coefficients can be difficult to achieve. To realize difficult-to-achieve transmission coefficients, one may need to conduct an extensive optimization, which complicates the design process. One may also need to employ some resonant metasurface element designs, which limits the metasurface's bandwidth. In our proposed work, the aggressive discretization level of only three unit cells per period allows a lower phase coverage requirement of only 240$^\circ$, which is almost realized with a rough parameter sweep, as shown in Fig.~\ref{Discussion}(a). Therefore, aggressive discretization in metasurface design can help one avoid using resonant unit cells and lead to broadband performance.

We compare the bandwidth (and efficiency) performance between discretized metasurface works of our group and some proposed metagrating designs. In the case of anomalous reflection with two propagating diffraction modes, a metagrating can be designed to realize efficient anomalous reflection~\cite{32,33}. However, the bandwidth is limited in order to achieve near-unity efficiency. One can achieve a wider bandwidth with compromised efficiency or more complicated design~\cite{32}. In~\cite{33}, the authors proposed a metagrating design with a bandwidth of 20.7\%, in which the anomalous reflection efficiency is higher than 90\%. Our group has proposed an aggressively discrete metasurface which can realize anomalous reflection with near-perfect efficiency~\cite{41}. The bandwidth with anomalous reflection power efficiency more than 90\% is 25.7\%, which is wider than the bandwidth presented in~\cite{33}. As for refraction, compared to reflection case with the same $k$-space operation, the number of propagating diffraction modes is doubled, resulting in worse anomalous refraction efficiency and/or bandwidth performance. In~\cite{35}, the authors proposed a metagrating which can realize anomalous refraction with total anomalous refraction power efficiency of 89\% and a 3-dB power efficiency bandwidth of 7.5\%. Our proposed design achieves a slightly lower efficiency of 82\% (anomalous refraction with normal incidence) but a wider 3-dB power efficiency bandwidth of 11\%. Importantly, the smallest feature size of our proposed design is about 5 times larger than that of~\cite{35}. Hence anomalous refraction is achieved with a relaxed feature size tolerance, leading to robust implementations amenable to straightforward frequency up-scaling.

While we have shown that our proposed metasurface can realize diffraction mode circulation with three anomalous refraction cases, it can also realize efficient anomalous refraction upon incidences with incident angles other than $-$45$^\circ$, 0$^\circ$ and 45$^\circ$. Fig.~\ref{Discussion}(b) shows the refraction angle ($\theta_t$) and power efficiency of the proposed metasurface as a function of the incidence angle ($\theta_i$). The refraction angle is the angle of the diffraction mode with non-zero magnitude within the propagation range of the transmission spectrum. From the $k$-space operation we can see that, for $-k_0<k_i<k_0-k_g$ ($-90^\circ<\theta_i<17^\circ$), the $1^\textrm{st}$ diffraction mode (with tangential wave number $k_1=k_i+k_g$) will be the only one with non-zero magnitude within the propagation range of the transmission spectrum; for $k_0-k_g<k_i<2k_g-k_0$ ($17^\circ<\theta_i<24.5^\circ$), there will be no diffraction mode with non-zero magnitude within the propagation range of the transmission spectrum; for $2k_g-k_0<k_i<k_0$ ($24.5^\circ<\theta_i<90^\circ$), the $-2^\textrm{nd}$ diffraction mode (with tangential wave number $k_{-2}=k_i-2k_g$) will be the only one with non-zero magnitude within the propagation range of the transmission spectrum. Then we use (3) to calculate the refraction angles. From Fig.~\ref{Discussion}(b) we can see that, when the incidence and refraction angles are both within 60$^\circ$ from normal, the power efficiency is more than 70\%. To further prove the design formulation, we show the cases of anomalous refraction of our proposed metasurface with $\theta_i$ = $-$50$^\circ$, $-$20$^\circ$ and 50$^\circ$ respectively. Figs.~\ref{Discussion}(c) and (d) are the transmission $k$-space spectrums and the corresponding transmission schematics of our proposed metasurface upon incidences with $\theta_i$ = $-$50$^\circ$, $-$20$^\circ$ and 50$^\circ$ respectively. From the $k$-space diagrams we can determine the refraction angles in the three cases to be $-$3.4$^\circ$, 21.4$^\circ$ and $-$40.4$^\circ$ respectively. Fig.~\ref{Discussion}(e) shows the simulated electric field magnitude distributions of our proposed metasurface upon Gaussian-beam illumination with incident angles of $-$50$^\circ$, $-$20$^\circ$ and 50$^\circ$ respectively. In each case, efficient anomalous refraction has been achieved with refraction angles which agree well with the $k$-space operation results.

\section{Conclusion}
We have shown, theoretically and experimentally, that aggressive discretization can benefit a metasurface design with simplicity and good performance. The aggressive discretization in metasurface design leads to much larger element size and lower requirement for phase/impedance coverage. This allows us to achieve efficient EM wave manipulation using a metasurface with very simple structure which relaxes fabrication tolerances. Besides, the decreased phase/impedance coverage requirement in aggressively discretized metasurface design allows one to avoid the use of unit cells with compromised properties or with strong resonance, resulting in efficient and broadband performance. Moreover, a discrete metasurface can realize diffraction mode circulation effect which cannot be realized by a finely discretized metasurface. Here we designed and fabricated a discrete transmissive Huygens’ metasurface with a simple structure --- it  features rectangular microstrip patterns, with the critical feature size 28-fold increased compared with a finely-discretized anomalous refraction Huygens' metasurface. It can realize efficient anomalous refraction by deflecting a normal incidence to 45$^\circ$. Meanwhile, due to its aggressively discretized design, the proposed metasurface can also realize efficient refraction of plane waves from $-$45$^\circ$ and 45$^\circ$ to 0$^\circ$ and $-$45$^\circ$ respectively, thus realizing diffraction mode circulation. Our proposed metasurface has a good performance with more than 80\% power efficiency and 11\% 3-dB power efficiency bandwidth. The concept of using a discrete metasurface to realize mode circulation effect can also be applied to guided waves and multi-channel communication systems.


%





\ifCLASSOPTIONcaptionsoff
  \newpage
\fi



%



\bibliographystyle{IEEEtran}

\begin{thebibliography}{10}

\bibitem{27}
Elena Abdo-S{\'a}nchez, Michael Chen, Ariel Epstein, and George~V
  Eleftheriades.
\newblock A leaky-wave antenna with controlled radiation using a bianisotropic
  {Huygens’} metasurface.
\newblock {\em IEEE Trans. Antennas Propag.}, 67(1):108--120, 2018.

\bibitem{7}
V.~S. Asadchy, Ana D{\'\i}az-Rubio, S.~N. Tcvetkova, D-H Kwon, A~Elsakka,
  M~Albooyeh, and S.~A. Tretyakov.
\newblock Flat engineered multichannel reflectors.
\newblock {\em Phys. Rev. X}, 7(3):031046, 2017.

\bibitem{8}
Viktar~S Asadchy, Mohammad Albooyeh, Svetlana~N Tcvetkova, Ana D{\'\i}az-Rubio,
  Younes Ra'di, and S.~A. Tretyakov.
\newblock Perfect control of reflection and refraction using spatially
  dispersive metasurfaces.
\newblock {\em Phys. Rev. B}, 94(7):075142, 2016.

\bibitem{37}
Andrea Casolaro, Alessandro Toscano, Andrea Al{\`u}, and Filiberto Bilotti.
\newblock Dynamic beam steering with reconfigurable metagratings.
\newblock {\em IEEE Trans. Antennas Propag.}, 68(3):1542--1552, 2019.

\bibitem{24}
Michael Chen, Elena Abdo-S{\'a}nchez, Ariel Epstein, and George~V
  Eleftheriades.
\newblock Theory, design, and experimental verification of a reflectionless
  bianisotropic {Huygens}' metasurface for wide-angle refraction.
\newblock {\em Phys. Rev. B}, 97(12):125433, 2018.

\bibitem{30}
Michael Chen and George~V Eleftheriades.
\newblock Omega-bianisotropic wire-loop {Huygens’} metasurface for
  reflectionless wide-angle refraction.
\newblock {\em IEEE Trans. Antennas Propag.}, 68(3):1477--1490, 2019.

\bibitem{31}
Michael Chen, Ariel Epstein, and George~V Eleftheriades.
\newblock Design and experimental verification of a passive {Huygens’}
  metasurface lens for gain enhancement of frequency-scanning slotted-waveguide
  antennas.
\newblock {\em IEEE Trans. Antennas Propag.}, 67(7):4678--4692, 2019.

\bibitem{22}
Francisco~S Cuesta, Ihar~A Faniayeu, Viktar~S Asadchy, and Sergei~A Tretyakov.
\newblock Planar broadband {Huygens’} metasurfaces for wave manipulations.
\newblock {\em IEEE Trans. Antennas Propag.}, 66(12):7117--7127, 2018.

\bibitem{28}
Manuel Decker, Isabelle Staude, Matthias Falkner, Jason Dominguez, Dragomir~N
  Neshev, Igal Brener, Thomas Pertsch, and Yuri~S Kivshar.
\newblock High-efficiency dielectric {Huygens’} surfaces.
\newblock {\em Adv. Opt. Mater.}, 3(6):813--820, 2015.

\bibitem{14}
Ayman~H Dorrah and George~V Eleftheriades.
\newblock Peripherally excited phased array architecture for beam steering with
  reduced number of active elements.
\newblock {\em IEEE Trans. Antennas Propag.}, 68(3):1249--1260, 2019.

\bibitem{29}
Ariel Epstein and George~V Eleftheriades.
\newblock Arbitrary power-conserving field transformations with passive
  lossless omega-type bianisotropic metasurfaces.
\newblock {\em IEEE Trans. Antennas Propag.}, 64(9):3880--3895, 2016.

\bibitem{19}
Ariel Epstein and George~V Eleftheriades.
\newblock Huygens’ metasurfaces via the equivalence principle: Design and
  applications.
\newblock {\em J. Opt. Soc. Am. B}, 33(2):A31--A50, 2016.

\bibitem{26}
Ariel Epstein, Joseph P~S Wong, and George~V Eleftheriades.
\newblock Cavity-excited {Huygens’} metasurface antennas for near-unity
  aperture illumination efficiency from arbitrarily large apertures.
\newblock {\em Nat. Commun.}, 7(1):1--10, 2016.

\bibitem{9}
Nasim~Mohammadi Estakhri and Andrea Alu.
\newblock Wave-front transformation with gradient metasurfaces.
\newblock {\em Phys. Rev. X}, 6(4):041008, 2016.

\bibitem{6}
Stanislav~B Glybovski, Sergei~A Tretyakov, Pavel~A Belov, Yuri~S Kivshar, and
  Constantin~R Simovski.
\newblock Metasurfaces: From microwaves to visible.
\newblock {\em Phys. Rep.}, 634:1--72, 2016.

\bibitem{1}
Christopher~L Holloway, Edward~F Kuester, Joshua~A Gordon, John O'Hara, Jim
  Booth, and David~R Smith.
\newblock An overview of the theory and applications of metasurfaces: The
  two-dimensional equivalents of metamaterials.
\newblock {\em IEEE Antennas Propag. Mag.}, 54(2):10--35, 2012.

\bibitem{2}
Christopher~L Holloway, Mohamed~A Mohamed, Edward~F Kuester, and Andrew
  Dienstfrey.
\newblock Reflection and transmission properties of a metafilm: With an
  application to a controllable surface composed of resonant particles.
\newblock {\em IEEE Trans. Electromagn. Compat.}, 47(4):853--865, 2005.

\bibitem{38}
Jiaqi Jiang, David Sell, Stephan Hoyer, Jason Hickey, Jianji Yang, and
  Jonathan~A Fan.
\newblock Free-form diffractive metagrating design based on generative
  adversarial networks.
\newblock {\em ACS Nano}, 13(8):8872--8878, 2019.

\bibitem{3}
Alexander~V Kildishev, Alexandra Boltasseva, and Vladimir~M Shalaev.
\newblock Planar photonics with metasurfaces.
\newblock {\em Science}, 339(6125), 2013.

\bibitem{21}
Minseok Kim, Alex M~H Wong, and George~V Eleftheriades.
\newblock Optical {Huygens’} metasurfaces with independent control of the
  magnitude and phase of the local reflection coefficients.
\newblock {\em Phys. Rev. X}, 4(4):041042, 2014.

\bibitem{lavigne2018susceptibility}
Guillaume Lavigne, Karim Achouri, Viktar~S Asadchy, Sergei~A Tretyakov, and
  Christophe Caloz.
\newblock Susceptibility derivation and experimental demonstration of
  refracting metasurfaces without spurious diffraction.
\newblock {\em IEEE Trans. Antennas Propag.}, 66(3):1321--1330, 2018.

\bibitem{13}
Aidin Mehdipour, Joseph~W Wong, and George~V Eleftheriades.
\newblock Beam-squinting reduction of leaky-wave antennas using huygens
  metasurfaces.
\newblock {\em IEEE Trans. Antennas Propag.}, 63(3):978--992, 2015.

\bibitem{15}
Kayode~Adedotun Oyesina and Alex M~H Wong.
\newblock Metasurface-enabled cavity antenna: Beam steering with dramatically
  reduced fed elements.
\newblock {\em IEEE Antennas Wireless Propag. Lett.}, 19(4):616--620, 2020.

\bibitem{20}
Carl Pfeiffer, Naresh~K Emani, Amr~M Shaltout, Alexandra Boltasseva, Vladimir~M
  Shalaev, and Anthony Grbic.
\newblock Efficient light bending with isotropic metamaterial {Huygens’}
  surfaces.
\newblock {\em Nano Lett.}, 14(5):2491--2497, 2014.

\bibitem{18}
Carl Pfeiffer and Anthony Grbic.
\newblock Metamaterial {Huygens}’ surfaces: Tailoring wave fronts with
  reflectionless sheets.
\newblock {\em Phys. Rev. Lett.}, 110(19):197401, 2013.

\bibitem{11}
Carl Pfeiffer and Anthony Grbic.
\newblock Bianisotropic metasurfaces for optimal polarization control: Analysis
  and synthesis.
\newblock {\em Phys. Rev. Appl.}, 2(4):044011, 2014.

\bibitem{12}
Carl Pfeiffer, Cheng Zhang, Vishva Ray, L~Jay Guo, and Anthony Grbic.
\newblock Polarization rotation with ultra-thin bianisotropic metasurfaces.
\newblock {\em Optica}, 3(4):427--432, 2016.

\bibitem{42}
Chu Qi and Alex M~H Wong.
\newblock A coarsely discretized {Huygens'} metasurface for anomalous
  transmission.
\newblock In {\em 2019 IEEE Asia-Pacific Microwave Conference (APMC)}, pages
  935--937. IEEE, 2019.

\bibitem{33}
Oshri Rabinovich and Ariel Epstein.
\newblock Analytical design of printed circuit board (pcb) metagratings for
  perfect anomalous reflection.
\newblock {\em IEEE Trans. Antennas Propag.}, 66(8):4086--4095, 2018.

\bibitem{35}
Oshri Rabinovich and Ariel Epstein.
\newblock Arbitrary diffraction engineering with multilayered multielement
  metagratings.
\newblock {\em IEEE Trans. Antennas Propag.}, 68(3):1553--1568, 2019.

\bibitem{34}
Oshri Rabinovich, Ilan Kaplon, Jochanan Reis, and Ariel Epstein.
\newblock Experimental demonstration and in-depth investigation of analytically
  designed anomalous reflection metagratings.
\newblock {\em Phys. Rev. B}, 99(12):125101, 2019.

\bibitem{32}
Younes Ra’di, Dimitrios~L Sounas, and Andrea Al{\`u}.
\newblock Metagratings: Beyond the limits of graded metasurfaces for wave front
  control.
\newblock {\em Phys. Rev. Lett.}, 119(6):067404, 2017.

\bibitem{39}
David Sell, Jianji Yang, Sage Doshay, Rui Yang, and Jonathan~A Fan.
\newblock Large-angle, multifunctional metagratings based on freeform multimode
  geometries.
\newblock {\em Nano Lett.}, 17(6):3752--3757, 2017.

\bibitem{17}
Michael Selvanayagam and George~V Eleftheriades.
\newblock Discontinuous electromagnetic fields using orthogonal electric and
  magnetic currents for wavefront manipulation.
\newblock {\em Opt. Express}, 21(12):14409--14429, 2013.

\bibitem{40}
Alex M~H Wong, Philip Christian, and George~V Eleftheriades.
\newblock Binary {Huygens’} metasurfaces: Experimental demonstration of
  simple and efficient near-grazing retroreflectors for {TE} and {TM}
  polarizations.
\newblock {\em IEEE Trans. Antennas Propag.}, 66(6):2892--2903, 2018.

\bibitem{41}
Alex M~H Wong and George~V Eleftheriades.
\newblock Perfect anomalous reflection with a bipartite {Huygens’}
  metasurface.
\newblock {\em Phys. Rev. X}, 8(1):011036, 2018.

\bibitem{25}
Alex M~H Wong and George~V Eleftheriades.
\newblock Active {Huygens'} box: Arbitrary electromagnetic wave generation with
  an electronically controlled metasurface.
\newblock {\em IEEE Trans. Antennas Propag.}, 69(3):1455--1468, 2021.

\bibitem{10}
Joseph P~S Wong, Ariel Epstein, and George~V Eleftheriades.
\newblock Reflectionless wide-angle refracting metasurfaces.
\newblock {\em IEEE Antennas Wireless Propag. Lett.}, 15:1293--1296, 2015.

\bibitem{16}
Joseph P~S Wong, Michael Selvanayagam, and George~V Eleftheriades.
\newblock Design of unit cells and demonstration of methods for synthesizing
  huygens metasurfaces.
\newblock {\em Photonics Nanostruct}, 12(4):360--375, 2014.

\bibitem{23}
Joseph P~S Wong, Michael Selvanayagam, and George~V Eleftheriades.
\newblock Polarization considerations for scalar huygens metasurfaces and
  characterization for 2-d refraction.
\newblock {\em IEEE Trans. Microw. Theory Tech.}, 63(3):913--924, 2015.

\bibitem{4}
Nanfang Yu and Federico Capasso.
\newblock Flat optics with designer metasurfaces.
\newblock {\em Nat. Mater.}, 13(2):139--150, 2014.

\bibitem{5}
Nanfang Yu, Patrice Genevet, Mikhail~A Kats, Francesco Aieta, Jean-Philippe
  Tetienne, Federico Capasso, and Zeno Gaburro.
\newblock Light propagation with phase discontinuities: Generalized laws of
  reflection and refraction.
\newblock {\em Science}, 334(6054):333--337, 2011.

\bibitem{36}
Ziying Zhang, Ming Kang, Xueqian Zhang, Xi~Feng, Yuehong Xu, Xieyu Chen,
  Huifang Zhang, Quan Xu, Zhen Tian, Weili Zhang, et~al.
\newblock Coherent perfect diffraction in metagratings.
\newblock {\em Adv. Mater.}, 32(36):2002341, 2020.

\end{thebibliography}

%








\end{document}